\documentclass[aps,twocolumn,showpacs,superscriptaddress,groupedaddress]{revtex4}  % for double-spaced preprint
\usepackage{graphicx}  % needed for figures
\usepackage{subfigure}
\usepackage{dcolumn}   % needed for some tables
\usepackage{bm}        % for math
\usepackage{amssymb,amsmath,amsopn,amsfonts}
\usepackage{colordvi}
\usepackage{color}
\usepackage[none]{hyphenat}
\begin{document}

% The following information is for internal review, please remove them for submission
\widetext
%\leftline{Version xx as of \today}
%\leftline{Primary authors: Joe E. Physics}
%\leftline{To be submitted to PRL.}
%\leftline{Comment to {\tt d0-run2eb-nnn@fnal.gov} by xxx, yyy}
%\centerline{\em D\O\ INTERNAL DOCUMENT -- NOT FOR PUBLIC DISTRIBUTION}

% the following line is for submission, including submission to the arXiv!!
%\hspace{5.2in} \mbox{Fermilab-Pub-04/xxx-E}

\title{Recursively Adaptive Quantum State Tomography: Theory and Two-qubit Experiment}
%\input author_list.tex       % D0 authors (remove the first 3 lines
                             % of this file prior to submission, they
                             % contain a time stamp for the authorlist)
                             % (includes institutions and visitors)
%\author{Bo Qi$^{1}$, Zhibo Hou$^{2}$, Li Li$^{2}$, Daoyi Dong$^{3}$, Guoyong Xiang$^{2,\star}$, Guangcan Guo$^{2}$\\
%$^{1}$\textit{Key Laboratory of Systems and Control, ISS, and National Center for Mathematics and Interdisciplinary Sciences, Academy of Mathematics and Systems Science, CAS, Beijing 100190, P. R. China}\\
%$^{2}$\textit{Key Laboratory of Quantum Information,University of Science and Technology of China, CAS, Hefei 230026, P. R. China}\\
%$^{3}$\textit{School of Engineering and Information Technology, University of New South Wales at the Australian Defence Force Academy, Canberra, ACT 2600, Australia}\\
%$^\star$\textit{email: gyxiang@ustc.edu.cn}}
\author{Bo Qi}
\thanks{These authors contributed equally to this work.}
\affiliation{Key Laboratory of Systems and Control, ISS, and National Center for Mathematics and Interdisciplinary Sciences, Academy of Mathematics and Systems Science, CAS, Beijing 100190, P. R. China}
\affiliation{Centre for Quantum Computation and Communication Technology and Centre for Quantum Dynamics, Griffith  University,  Brisbane,  Queensland  4111,  Australia}
\author{Zhibo Hou}
\thanks{These authors contributed equally to this work.}
\affiliation{Key Laboratory of Quantum Information, University of Science and Technology of China, CAS, Hefei, 230026, People's Republic of China}
\affiliation{Synergetic Innovation Center of Quantum Information and Quantum Physics, University of Science and Technology of China, Hefei, Anhui 230026, People's Republic of China}
\author{Yuanlong Wang}
\affiliation{School of Engineering and Information Technology, University of New South Wales, Canberra, ACT 2600, Australia}
\author{Daoyi Dong}
\affiliation{School of Engineering and Information Technology, University of New South Wales, Canberra, ACT 2600, Australia}
\author{Han-Sen Zhong}
\affiliation{Key Laboratory of Quantum Information, University of Science and Technology of China, CAS, Hefei, 230026, People's Republic of China}
\affiliation{Synergetic Innovation Center of Quantum Information and Quantum Physics, University of Science and Technology of China, Hefei, Anhui 230026, People's Republic of China}
\author{Li Li}
\affiliation{Centre for Quantum Computation and Communication Technology and Centre for Quantum Dynamics, Griffith  University,  Brisbane,  Queensland  4111,  Australia}
\author{Guo-Yong Xiang}
%\email{gyxiang@ustc.edu.cn}
\thanks{gyxiang@ustc.edu.cn}
\affiliation{Key Laboratory of Quantum Information, University of Science and Technology of China, CAS, Hefei, 230026, People's Republic of China}
\affiliation{Synergetic Innovation Center of Quantum Information and Quantum Physics, University of Science and Technology of China, Hefei, Anhui 230026, People's Republic of China}
\author{Howard M. Wiseman}
\affiliation{Centre for Quantum Computation and Communication Technology and Centre for Quantum Dynamics, Griffith  University,  Brisbane,  Queensland  4111,  Australia}
\author{Chuan-Feng Li}
\affiliation{Key Laboratory of Quantum Information, University of Science and Technology of China, CAS, Hefei, 230026, People's Republic of China}
\affiliation{Synergetic Innovation Center of Quantum Information and Quantum Physics, University of Science and Technology of China, Hefei, Anhui 230026, People's Republic of China}
\author{Guang-Can Guo}
\affiliation{Key Laboratory of Quantum Information, University of Science and Technology of China, CAS, Hefei, 230026, People's Republic of China}
\affiliation{Synergetic Innovation Center of Quantum Information and Quantum Physics, University of Science and Technology of China, Hefei, Anhui 230026, People's Republic of China}

\date{\today}

\begin{abstract}
%We present a new adaptive and recursive quantum state tomography  protocol using adaptive linear regression estimation (ALRE). It is demonstrated numerically that our ALRE, even if wherein the measurement strategy is limited to the simplest tensor product measurements, outperforms the tomography protocols using mutually unbiased bases and some other adaptive strategies wherein non-local measurements are involved. Furthermore, if all measurement strategies can be performed, the infidelity of our ALRE can be reduced to beat the Gill-Massar bound  for a wide range of quantum states with modest number of copies. Our ALRE protocol is particularly efficient for maximally entangled states, which are important resources in quantum information.
Adaptive techniques have important potential for wide applications in enhancing precision of quantum parameter estimation. We present a recursively adaptive quantum state tomography (RAQST) protocol for finite dimensional quantum systems and experimentally implement the adaptive tomography protocol on two-qubit systems. In this RAQST protocol, an adaptive measurement strategy and a recursive linear regression estimation algorithm are performed. Numerical results show that our RAQST protocol can outperform the tomography protocols using mutually unbiased bases (MUB) and the two-stage MUB adaptive strategy even with the simplest product measurements. When nonlocal measurements are available, our RAQST can beat the Gill-Massar bound for a wide range of quantum states with a modest number of copies. We use only the simplest product measurements to implement two-qubit tomography experiments. In the experiments, we use error-compensation techniques to tackle systematic error due to misalignments and imperfection of wave plates, and achieve about 100-fold reduction of the systematic error. The experimental results demonstrate that the improvement of RAQST over nonadaptive tomography is significant for states with a high level of purity. Our results also show that this recursively adaptive tomography method is particularly effective for the reconstruction of maximally entangled states, which are important resources in quantum information.
\end{abstract}

%\pacs{03.65.Wj, 02.50.-r, 03.67.-a}

%\pacs{03.67.-a, 03.65.Wj, 06.20.Dk, 03.65.Ta}
%03.67.-a: quantum information
%03.65.Wj: quantum tomography, state reconstruction
%03.65.-w: quantum mechanics
%03.65.Ta: Foundations of quantum mechanics

%06.20.Dk: Measurement and error theory
%42.50.Dv: Quantum state engineering and measurements
%42.25.Ja Polarization

%03.67.Mn: Entanglement production, characterization and manipulation
%03.65.Ud: Entanglement and quantum nonlocality
%(e.g. EPR paradox, Bell's inequalities, GHZ states, etc.)
%(for entanglement production in quantum information, see 03.67.Mn;

%02.10.De: algebraic structure

\maketitle

\section{\label{sec:intro}introduction}
% sections are not used for PRL papers
One of the central problems in quantum science and technology is the estimation of an unknown quantum state \cite{Niel00quantum}.
Quantum state tomography as the procedure of experimentally determining an unknown quantum state has become a standard technology for verification and benchmarking of quantum devices \cite{GillM00,Jame01measurement,Pari04quantum,Toth10permutationally,GrosLFB10,Reha10operational,Cram10efficient,ChriR12,Weit12experimental,Jeff12procedure,Qi13quantum,Taka13precision,Klim13optimal,Salv13full,Mira14optimal}. Two key tasks in quantum state tomography are data acquisition and data analysis. The aim of data acquisition is to devise appropriate measurement strategies to acquire information for reconstructing the quantum state. Then in the step of data analysis, the acquired data is associated with an estimate of the unknown quantum state using an estimation algorithm.

%The aim of data acquisition is to devise appropriate measurement strategies that are efficient to acquire information for reconstructing the quantum state. There have been many results concerning the optimal measurement design for quantum state tomography \cite{Woot89optimal,Burg08choice,BisiCDF09P,Nunn10optimal,Adam10improving}.  For instance, improved state estimation can be achieved by taking advantage of mutually unbiased bases as compared to standard measurement strategies \cite{Adam10improving}. However, mutually unbiased measurements are difficult to realize experimentally, since they involve non-local measurements. How to efficiently acquire information of an unknown quantum state with measurements that are simple to realize experimentally remains open.

In order to enhance the efficiency in data acquisition, it is desired to develop optimal measurement strategies for collecting data. However, an optimal measurement strategy, which is only known for a few special cases~\cite{GillM00,Hole82book,Haya05book,Zhu12quantum,Geng14experimental}, depends on the state to be reconstructed. To circumvent this issue, many kinds of fixed sets of measurement bases are designed to be optimal either in terms of the average over a certain quantum state space \cite{Woot89optimal,Burg08choice,BisiCDF09P,Nunn10optimal,Adam10improving} or in terms of the worst case in the quantum state space \cite{Qi13quantum}. For instance, improved state estimation can be achieved by taking advantage of mutually unbiased bases (MUB) \cite{Woot89optimal,Durt10on,Adam10improving} and symmetric informationally complete positive operator-valued measures (SIC-POVM) \cite{Rene04symmetric,Bent15experimental}. For multi-partite quantum systems, MUB and SIC-POVM are difficult to experimentally realize since they involve nonlocal measurements. How to efficiently acquire information of an unknown quantum state using simple measurements that are easy to realize experimentally remains open.

For data analysis in tomography, although many methods, such as maximum-likelihood estimation \cite{Pari04quantum,Blum10H,TeoZER11,Teo12incomplete,Smol12efficient}, Bayesian mean estimation \cite{Husz12adaptive,Blum10O}, least-squared inversion \cite{Opat97least}, have been used to reconstruct the quantum state, this task can be  computationally intensive, and may take even more time than the  experiments themselves. It has been reported in \cite{GrosLFB10} that using the maximum-likelihood method to reconstruct eight-qubit took weeks of computation. Therefore, the development of an efficient data analysis algorithm is also a critical issue in quantum state tomography \cite{Mahl13adaptive,Qi13quantum}. In \cite{Qi13quantum}, a recursive linear regression estimation algorithm was presented which is much more computationally efficient in the sense that it can greatly save the cost of computation as compared to the maximum-likelihood method with only a small amount of accuracy sacrificed.

For a given number of copies of the system, in order to improve the tomography accuracy by better tomographic measurements, a natural idea is to develop an adaptive tomography protocol where the measurement can be adaptively optimized based on data collected so far. Adaptive measurements have shown more powerful capability than nonadaptive measurements in quantum phase estimation \cite{Wise95adaptive,Higg07entanglement,Xian11entanglement}, phase tracking \cite{Yone12quantum}, quantum state discrimination \cite{Acin05multiple,Higg09mixed}, and Hamiltonian estimation \cite{Serg11characterization,Ferr13how}. Actually, adaptivity has been proposed for quantum state tomography in various contexts \cite{GillM00,Husz12adaptive,BagaBGM06S,OkamIOY12,Sugi12adaptive,Mahl13adaptive,Krav13experimental,Lerc14adaptive}. For example, the results on one qubit have demonstrated that adaptive quantum state tomography can improve the accuracy quadratically considering the infidelity index \cite{Mahl13adaptive}. However, when generalizing their results to $n$-qubit systems, the adaptive tomography protocol will involve nonlocal measurements which are hard to realize in experiments.

In this paper, we combine the computational efficiency of the recursive technique of \cite{Qi13quantum} with a new adaptive protocol that does not necessarily require nonlocal measurement to present a new recursively adaptive quantum state tomography (RAQST) protocol. In our RAQST protocol, no prior assumption is made on the state to be reconstructed. The state estimate is recursively updated based on the current estimate and the new measurement data. Thus, we do not have to combine all the historical information with the new acquired data to update the estimate as the maximum-likelihood method.  Thanks to the simple recursive estimation procedure, we can obtain the estimate state in a realtime way, and using the estimate we can adaptively optimize the measurement strategies to be performed in the following step. In our RAQST protocol, the measurement to be performed at each step is optimized upon the corresponding admissible measurement set determined by the experimental conditions.

It is first demonstrated numerically that our RAQST even with the simplest product measurements can outperform the tomography protocols  using MUBs and the two-stage MUB adaptive strategy. For maximally entangled states, the infidelity can even be reduced to beat the Gill-Massar bound which is a quantum Cram\'{e}r-Rao inequality \cite{GillM00}. Moreover, if nonlocal measurements are available, with our RAQST the infidelity can be further reduced. For a wide range of quantum states, the infidelity of our RAQST can be reduced to beat the Gill-Massar bound with a modest number of copies. We perform the two-qubit state tomography experiments using only the simplest product measurements, and the experimental results demonstrate that the improvement of our RAQST over nonadaptive tomography  is significant for states with a high level of purity. This limit (very high purity)
is the one relevant for most forms of quantum information processing.
%The paper is organized as follows: first, we give the recursively adaptive linear regression estimation %method for the quantum state tomography in Section II; then we present the numerical results in Section %III; in Section IV, we introduce the experimental setup and present the experimental results; the %summary and perspectives are given in Section V.
%\textbf{Adaptive linear regression estimation}
\section{\label{sec:ALRE}Recursively Adaptive linear regression estimation}

A linear regression estimation (LRE) method for quantum state tomography was proposed in \cite{Qi13quantum}, and the results have shown that the LRE approach has much lower computational complexity than the maximum-likelihood estimation method for quantum tomography. Here, we further develop this LRE method to present a recursively adaptive quantum state tomography protocol that can greatly improve the precision of tomography.

We first convert a quantum state tomography problem into a parameter estimation problem of a linear regression model. Consider a $d$-dimensional quantum system with Hilbert space $\mathcal{H}$. Let $\{\Omega_{i}\}^{d^2-1}_{i=1}$ denote a  set of Hermitian operators satisfying (i) $\textmd{Tr}(\Omega_{i})=0$ and  (ii) $\textmd{Tr}(\Omega_i\Omega_j)=\delta_{ij}$, where $\delta_{ij}$ is the Kronecker function. Using this set, the quantum state $\rho$ to be reconstructed can be parameterized as
\begin{equation}\label{rho}
\rho=\frac{I}{d}+\sum^{d^2-1}_{i=1}\theta_i\Omega_i,
\end{equation}
where $I$ is the identity matrix and $\theta_i=\textmd{Tr}(\rho\Omega_i)$. Let $\Theta=(\theta_1, \cdots, \theta_{d^2-1})^{T}$, where $T$ denotes the transpose operation.

A quantum measurement can be described by a positive operator-valued measure (POVM) $\{E_i\}_{i=1}^{M}$, which is a set of positive semidefinite matrices that sum to the identity, i.e., $E_i\geq 0$ and $\sum_{i=1}^M E_i=I$. In quantum state tomography, different sets of POVMs should be appropriately combined to efficiently acquire information of the unknown quantum state. Let $\mathcal{M}={\underset{j=1}{\bigcup}}\mathcal{M}^{(j)}$ denote the admissible measurement set, which is
 a union of POVMs determined by the experimental conditions.  Each POVM is denoted as $\mathcal{M}^{(j)}=\{E_i^{(j)}\}^{M^{(j)}}_{i=1}$.
Using the set of $\{\Omega_{k}\}^{d^2-1}_{k=1}$, elements of the POVM can be parameterized as
\begin{equation*}\label{povm}
E_i^{(j)}=\gamma^{(j)}_{i,0}\frac{I}{d}+\sum^{d^2-1}_{k=1}\gamma^{(j)}_{i,k}\Omega_k,
\end{equation*}
where $\gamma^{(j)}_{i,0}=\textmd{Tr}(E_i^{(j)})$, and $\gamma^{(j)}_{i,k}=\textmd{Tr}(E_i^{(j)}\Omega_k)$.
Let $\Gamma^{(j)}_i=(\gamma^{(j)}_{i,1}, \cdots, \gamma^{(j)}_{i,d^2-1})^{T}.$ When we perform the POVM  $\mathcal{M}^{(j)}$ on copies of a system in state $\rho$, the probability that we observe the result $m$ is given by
\begin{equation}\label{averageequation}
p(m|\mathcal{M}^{(j)})=\textmd{Tr}(E^{(j)}_m\rho)
=\gamma^{(j)}_{m,0}/d+\Theta^{T}{\Gamma^{(j)}_m}.
\end{equation}

Assume that the total number of experiments is $N$, and we perform a measurement described by $\mathcal{M}^{(j)}=\{E_i^{(j)}\}_{i=1}^{M^{(j)}}$ $n^{(j)}$ times. Let $n^{(j)}_m$ denote the number of the occurrence of the outcome $m$ from the $n^{(j)}$ measurement trials of $\mathcal{M}^{(j)}$. Let $\hat{p}(m|\mathcal{M}^{(j)})=n^{(j)}_m/n^{(j)}$, and $e_m^{(j)}=\hat{p}(m|\mathcal{M}^{(j)})-p(m|\mathcal{M}^{(j)})$. According to the central limit theorem, $e_m^{(j)}$ converges in distribution to a normal distribution with mean 0 and variance
$[p(m|\mathcal{M}^{(j)})-p^2(m|\mathcal{M}^{(j)})]/n^{(j)}$. Using (\ref{averageequation}), we have the linear regression equations for $m=1,\ \cdots,\ M^{(j)}$,
\begin{equation}\label{average2}
\hat{p}(m|\mathcal{M}^{(j)})=\gamma^{(j)}_{m,0}/d+\Theta^{T}{\Gamma^{(j)}_m}+e_m^{(j)}.
\end{equation}
Note that $\hat{p}(m|\mathcal{M}^{(j)})$, $\gamma^{(j)}_{m,0}/d$  and $\Gamma^{(j)}_m$ are all available, while $e_m^{(j)}$ may be considered as the observation noise. Hence, the problem of quantum state tomography is converted into the estimation of the unknown vector $\Theta$.

To give an estimate with a high level of accuracy, the basic idea of LRE is to find an estimate $\hat{\Theta}_t$ such that
\begin{widetext}
\begin{equation}\label{theta}
\hat{\Theta}_{t}=\underset{\hat{\Theta}}{\text{argmin}}
\sum_{k=1}^t\sum^{M^{(j_k)}}_{m=1}W^{(j_k)}_m[\hat{p}(m|\mathcal{M}^{(j_k)})-\gamma^{(j_k)}_{m,0}/d-
{\hat{\Theta}}^{T}{\Gamma^{(j_k)}_m}]^2.
\end{equation}
\end{widetext}
Here, $\mathcal{M}^{(j_k)}$ denotes the POVM $\mathcal{M}^{(j_k)}=\{E_m^{(j_k)}\}_{m=1}^{M^{(j_k)}}$ being performed at the $k$-th step. The notation
$W^{(j_k)}_m$ denotes the weight of the corresponding linear regression equation. In general, the smaller the variance of $e_m^{(j_k)}$ is, the more the information can be extracted by $E_m^{(j_k)}$. Therefore, the corresponding  weight of the  regression equation should be larger. A sound choice of $W^{(j_k)}_m$ is the estimate of the inverse of the variance of $e^{(j_k)}_m$, i.e., $W^{(j_k)}_m=n^{(j_k)}/[\hat{p}(m|\mathcal{M}^{(j_k)})-\hat{p}^2(m|\mathcal{M}^{(j_k)})]$.

A recursive LRE algorithm \cite{Qi13quantum} can be utilized to find the solution of $\hat{\Theta}_t$. For completeness, we present the recursive LRE algorithm in Appendix A. Its basic idea is that one only needs to store the best estimate state so far, and then update it recursively using a bunch of new measurement results  with a fixed setting.  This is quite different from the maximum-likelihood estimation method since there one has to combine all the historical information with the new collected data to update the estimate, which is quite computationally intensive. It has been demonstrated in Fig. 1 of \cite{Qi13quantum} that the recursive LRE tomography algorithm can greatly reduce the total cost of computation with only a small amount of accuracy sacrificed in comparison with the  maximum-likelihood estimation method.

As demonstrated in Appendix B, when the number of copies $N$ of the unknown quantum state becomes large, the only relevant measure of the quality of estimation becomes the mean squared error  matrix $E(\hat{\Theta}_t-\Theta)(\hat{\Theta}_t-\Theta)^{T}$. The mean squared error matrix depends upon the state $\rho$ (i.e., $\Theta$) to be reconstructed and the chosen POVMs. Thanks to the recursive algorithm, we can obtain the estimate of the state $\rho$ recursively, and then adaptively optimize the POVM measurements that should be performed. By doing so, the accuracy of the tomography can be greatly improved. The details of how to adaptively choose POVMs are presented in Appendix C.

Using the solution $\hat{\Theta}_{t}$ in (\ref{theta}) and the relationship in (\ref{rho}), we can obtain a Hermitian matrix $\hat{\mu}$ with $\textmd{Tr}\hat{\mu}=1$. However, $\hat{\mu}$ may have negative eigenvalues and be nonphysical due to the randomness of measurement results. In this work, the  physical estimate $\hat{\rho}$ is chosen to be the closest density matrix to $\hat{\mu}$ under the matrix 2-norm. In standard state reconstruction algorithms, this task is computationally intensive \cite{Smol12efficient}. However, we can employ the fast algorithm in \cite{Smol12efficient} with computational complexity $O(d^3)$ to solve this problem since we have a Hermitian estimate $\hat{\mu}$ with $\textmd{Tr}\hat{\mu}=1$.
It can be verified that pulling $\hat{\mu}$ back to a physical state can further reduce the mean squared error \cite{Taka13precision}.

%\textbf{Numerical results}
\section{\label{sec:Numerical results}Numerical results}

In this section we present the numerical results. First of all, we would like to stress two advantages of the recursive LRE method: (a) as we have demonstrated in \cite{Qi13quantum}, the recursive LRE method can greatly reduce the cost of computation in comparison with the maximum-likelihood method;
%with only a small amount of accuracy sacrificed;
(b) the recursive LRE algorithm is naturally suitable for optimizing measurements adaptively. The argument for the advantage (b) can be explained as follows. For state tomography the optimal measurements generally depend upon the state to be reconstructed. By utilizing the recursive LRE algorithm, we can obtain the estimate of the real state in a computationally efficient way. Using the state estimate, the measurements to be performed can be adaptively optimized. In the following, we perform numerical simulations of two-qubit tomography using only the recursive LRE method while with six different measurement strategies: (i) standard cube measurements \cite{Burg08choice}; (ii) mutually unbiased bases (MUB) measurements; (iii) MUB half-half \cite{Mahl13adaptive}; (iv) ``known basis" \cite{Mahl13adaptive}; (v) RAQST1: the admissible measurement set only contains the simplest product measurements; (vi) RAQST2: the admissible measurement set is not limited.

Each of the tomography protocols (iii)-(vi) consists of two stages. In the first stage, we all use the standard cube measurements.  For the MUB half-half,  we first perform standard cube measurements on  $N/2$ copies and obtain a preliminary estimate $\hat{\rho}_0$ via LRE, and then  measure the remaining half of copies so that one set of the bases is adaptively adjusted to diagonalize $\hat{\rho}_0$ and it together with another four sets of bases constitutes a complete set of MUB as proposed in \cite{Mahl13adaptive}.  As compared to the MUB half-half, for the ``known basis" \cite{Mahl13adaptive}, in the second stage, we perform a set of measurements so that one of the five bases of the MUB is the eigenbasis of the state to be reconstructed. Although it is impossible physically, this is a useful comparison.
For the RAQST, we first perform standard cube measurements on $N_1$ copies and obtain a preliminary estimate. Then we adaptively optimize the measurement to be performed at each iteration step upon the corresponding admissible measurement set (see  Appendix D). In RAQST1, the basic admissible measurement set is the standard cube measurement bases. At each iteration step, we add another set of product measurements obtained by solving a conditional extremum problem to the basic admissible measurement set (see Appendix E). In RAQST2, at each iteration step, the set of the eigenbases of the  current estimate state is also added into the admissible measurement set. Note that
the admissible measurement set in RAQST2 will involve nonlocal measurements in general if there are more than one particle. The details can be found in Appendix D.

For the RAQST, we need to specify $N_1$, which is the number of copies measured in the first stage, and the number $K$ of the iteration steps such that $N=N_1+K\cdot N_2$, where $N_2$ is the number of copies for each POVM in the second stage. In principle, the number $K$ of the iteration steps in the second stage may depend on the preliminary estimate in the first stage. For simplicity, in this work, we give  empirical formulas  depending only upon the total number $N$ of the copies. Note that in RAQST1 and RAQST2, the admissible measurement sets are different, and so are their empirical formulas.
For RAQST1, $N^{(1)}_1=N/(1.3+0.1\log_{10}{N})$,  $K^{(1)}=\lfloor \log_{10}{N}-1\rfloor$, and for RAQST2, $N^{(2)}_1=N(0.8-0.01\log_{10}{N})$,  $K^{(2)}=\lfloor 1.5\log_{10}{N}-2\rfloor$ where $\lfloor x\rfloor$ returns the maximum integer that is less than or equal to $x$. Obviously the formula for the resource distribution for RAQST2 applies only when $N$ is not too large.

We use Monte Carlo simulations to demonstrate the results. The  figure of merit is the particularly well-motivated quantum infidelity \cite{Mahl13adaptive}, $1-F(\rho, \hat{\rho})=1-\mathrm{Tr}^2(\sqrt{\sqrt{\rho}\hat{\rho}\sqrt{\rho}})$.
\begin{figure}[]
\centering
\subfigure[]{
\label{fig0:subfig:a}
\includegraphics[width=0.47\textwidth]{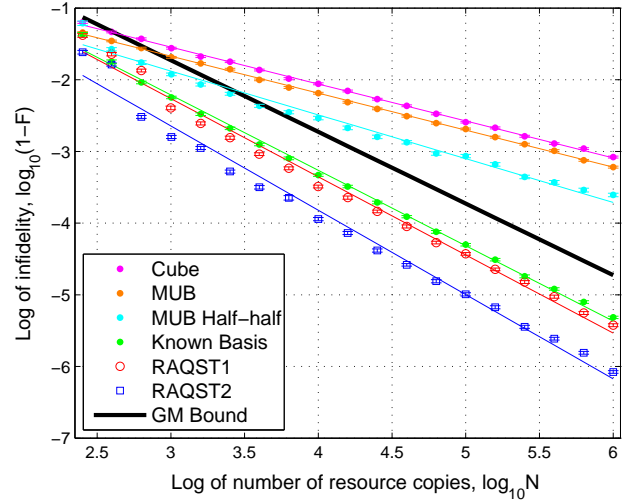}}
\subfigure[]{
\label{fig0:subfig:b}
\includegraphics[width=0.47\textwidth]{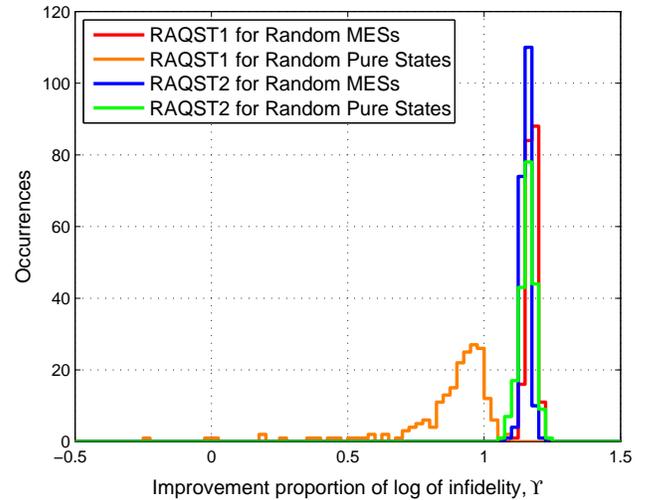}}
\caption{Performance of the RAQST protocol for pure states. (a) Infidelity versus $N$ for $\frac{|HV\rangle-|VH\rangle}{\sqrt2}$ with different tomography protocols. Each point is averaged over 100 realizations. Error bars are the standard deviation of the average. (b) Histogram of improvement proportion of infidelity for 200 randomly selected maximally entangled states (MESs) and 200 pure states when the total number of copies is $N=10^4$ for each random state. Each generated state is repeated through the RAQST protocol for 200 times.}
\label{fig0:subfig}
\end{figure}
Fig.~\ref{fig0:subfig:a} depicts average infidelity versus $N$ for the maximally entangled state $\frac{|HV\rangle-|VH\rangle}{\sqrt2}$. It can be seen that the average infidelity of the static tomography protocols (i.e., (i) and (ii)) versus $N$ is in the order of $O(1/\sqrt{N})$. However, the Gill-Massar bound \cite{GillM00} for the infidelity in two-qubit state tomography  is $\frac{75}{4N}$.  This can be obtained by combining the equations (5.29) and (A.8) in \cite{Zhu12quantum} (see Appendix~\ref{sec:append: GM bound}).  It is clearly seen that, as compared to the static tomography protocols  and the adaptive  MUB half-half, the average infidelity using our RAQST protocol can be reduced to beat the Gill-Massar bound even only with the simplest product measurements. Furthermore, if there is no limitation on the admissible measurement set, the RAQST2 can outperform the ``known basis" tomography, and the average infidelity of RAQST2 versus $N$ can be significantly reduced to the order of the Gill-Massar bound, i.e., $O(1/N)$.

Fig.~\ref{fig0:subfig:b} shows the histogram for RAQST over 200 randomly selected pure states and 200 maximally entangled states when the total number of copies is $N=10^4$ for each random state. Random pure states are created using the algorithm in \cite{Zycz94random}. Since all the maximally entangled states are equivalent under local unitary operations, they are randomly selected by applying randomly generated local unitary operators \cite{Misz12generating} on the same  maximally entangled states. We adopt the index $\Upsilon=\frac{C-A}{C-G} $ to evaluate the performance of our RAQST protocol. Here, $C$ and $A$ represent the $\log_{10}$ of the average infidelity between the corresponding estimate and the true state when the standard cube measurement bases and the RAQST are utilized, respectively, while $G$ is the Gill-Massar bound. Note that if $\Upsilon>0$, our adaptive protocol surpasses the standard measurement strategy, while if $\Upsilon>1$, our adaptive protocol beats the Gill-Massar bound. From Fig. \ref{fig0:subfig:b} we can see that our RAQST protocol is particularly effective for the class of  maximally entangled states which are important resources in quantum information.

\begin{figure}[]
\centering
\subfigure[]{
\label{fig:subfig:a}
\includegraphics[width=0.47\textwidth]{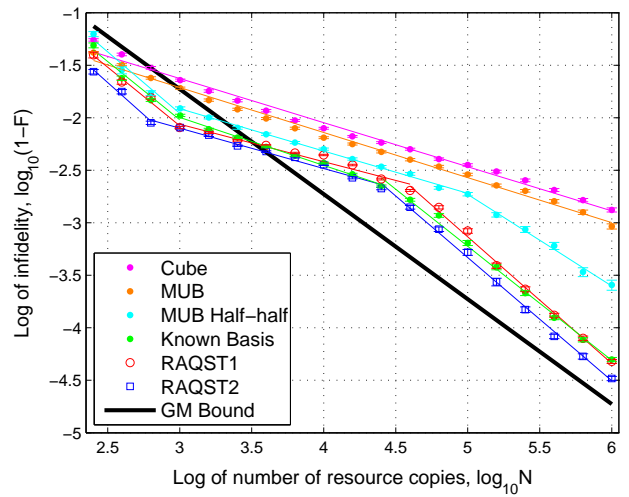}}
\subfigure[]{
\label{fig:subfig:b}
\includegraphics[width=0.47\textwidth]{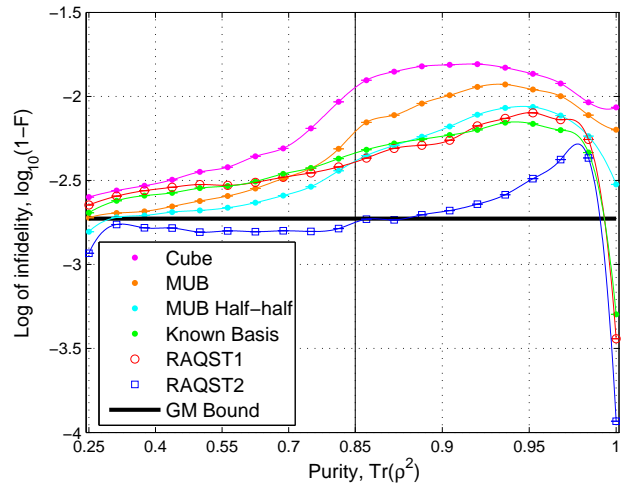}}
\caption{Performance of the RAQST protocol for mixed states. (a) Infidelity versus $N$ with different tomography protocols for state $\rho=0.997 \frac{(|HV\rangle-|VH\rangle)(\langle HV|-\langle VH|)}{2}+0.003 \frac{I}{4}$, which has purity Tr$(\rho^2)$=0.9955. Each point is averaged over 200 realizations. Error bars are the standard deviation of the average. (b) Infidelity versus different purity with different tomography methods for quantum states $\alpha \frac{(|HV\rangle-|VH\rangle)(\langle HV|-\langle VH|)}{2}+\beta \frac{I}{4}$, where $\alpha,\ \beta\geq0$ and $\alpha+\beta=1$. The total number of copies for each state is $N=10^4$. Each point is averaged over 1000 realizations. }
\label{fig:subfig}
\end{figure}

%Fig. \ref{fig:subfig:a} shows average infidelity versus $N$ for state $\rho=0.997 \frac{(|HV\rangle-|VH\rangle)(\langle HV|-\langle VH|)}{2}+0.003 \frac{I}{4}$, which has purity Tr$(\rho^2)$=0.9955. It can be seen that the infidelity is lower by using our ALRE1 with only tensor product measurements than those by using the static tomograph protocols (i), (ii),  and the adaptive MUB half-half (iii). The infidelity can be further reduced by using ALRE2.

Fig.~\ref{fig:subfig:a} depicts average infidelity versus $N$ for state $\rho=0.997 \frac{(|HV\rangle-|VH\rangle)(\langle HV|-\langle VH|)}{2}+0.003 \frac{I}{4}$, which has purity Tr$(\rho^2)$=0.9955. Note that there are kinks in the four curves corresponding to the four different adaptive protocols (iii)-(vi).  We can see that each of the four curves can be divided into three segments from left to right. In the first segment, the infidelity decreases quickly as $N$ increases until the infidelity is reduced to the order of the small eigenvalues of the state to be reconstructed, then the curves go into the second segment where the infidelity decreases slowly. After the infidelity is smaller than the smallest eigenvalues, the infidelity decreases quickly again as $N$ increases. This is because infidelity is hypersensitive to misestimation of small eigenvalues, as pointed out in \cite{Mahl13adaptive}. Hence, we must accurately estimate the eigenvalues that appear to be zero. When the infidelity is in the order of the smallest eigenvalues, it will be hard to estimate them accurately, so the decay rate of the infidelity will become slow. Once the infidelity decreases to be smaller than the smallest eigenvalues, we can estimate them more accurately as $N$ increases, and then the infidelity decreases quickly. It can be seen that our RAQST1 can beat the static tomography protocols and the adaptive MUB half-half protocol even with the simplest product measurements. The infidelity can be further reduced by using RAQST2, and when the total copies  $N\ge 10^{4.5}$, the infidelity can be reduced to $O(1/N)$.

Fig.~\ref{fig:subfig:b} shows average infidelity versus different purity when the total number of the copies for each state is $N=10^4$. The quantum states are $\alpha \frac{(|HV\rangle-|VH\rangle)(\langle HV|-\langle VH|)}{2}+\beta \frac{I}{4}$, where $\alpha,\ \beta\geq0$ and satisfy $\alpha+\beta=1.$
The results show that when the states have a high level of purity, our RAQST1 with the simplest product measurements can beat the MUB protocol. However, as the state becomes more mixed ($\textmd{Tr}(\rho^2)$ decreases), using MUB measurements for state tomography can do better than using the adaptive product measurements. This fact is due to the essential limit of product measurements on mixed states. As pointed out in \cite{GillM00}, nonlocal measurements on a mixed state can extract more information. Thus, to reconstruct mixed states, it is better to use nonlocal measurements, e.g., MUB measurements. It is also clear that the infidelity achieved by using RAQST2 is much lower than that using MUB, and can beat the Gill-Massar bound for a wide range of quantum states.

%\textbf{Experimental results}
\section{\label{sec:Experimental results}Experimental results}
In this section, we report the experimental results using our RAQST protocol for two-qubit quantum state tomography. Since it is hard to perform nonlocal measurements in real experiments, we only experimentally implement tomography protocols using (i) standard cube measurements and (v) RAQST1.

\begin{figure}
\center{\includegraphics[scale=0.55]{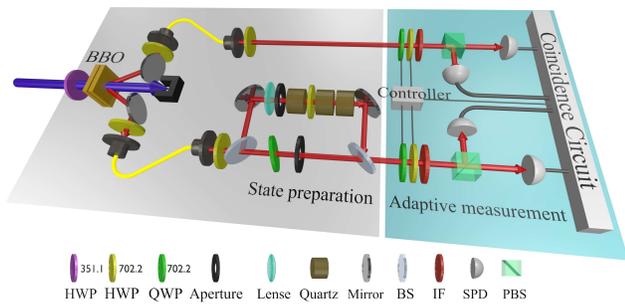}}% Here is how to import EPS art
\caption{\label{experimental_setup}(color online) {Experimental setup. The experimental setup can be divided into two modules: state preparation (gray) and adaptive measurement (light blue). In state preparation module, arbitrary Werner states in the form $\rho=p\frac{(|HV\rangle-|VH\rangle)(\langle HV|-\langle VH|)}{2}+(1-p)\frac{I}{4}$ can be generated. In adaptive measurement module, the two-photon product measurements are realized and can be adaptively changed by a Labview program. Key to components: HWP, half-wave plate; QWP, quarter-wave plate; BS, beam splitter; IF, interference filter; SPD, single photon detector; PBS, polarizing beam splitter.}}
\end{figure}

As shown in Fig.~\ref{experimental_setup}, the experimental setup includes two modules: state preparation (gray) and adaptive measurement (light blue). In the state preparation module, a pair of polarization-entangled photons with a central wavelength at $\lambda=$702.2 nm is first generated after the continuous Ar$^+$ laser at $351.1$ nm with diagonal polarization pumps a pair of type I phase-matched $\beta$-barium borate (BBO) crystals whose optic axes are normal to each other \cite{Kwia99Ultrabright}. The generation rate is about 3000 two-photon coincidence counts per second at a pump power of 60 mW. Half-wave plates at both ends of the two single mode fibers are used to control polarization. Then, one photon is either reflected by or transmits through a 50/50 beam splitter (BS). In the transmission path, a QWP is tilted to compensate the phase of the two-photon state for the generation of $\frac{|HV\rangle-|VH\rangle}{\sqrt{2}}$. In the reflected path, three 446 $\lambda$ quartz crystals and a half wave plate with 22.5$^\circ$ are used to dephase the two-photon state into a completely mixed state $I/4$. The ratio of the two states mixed at the output port of the second BS can be changed by the two adjustable apertures for the generation of arbitrary Werner state in the form $\rho=p\frac{(|HV\rangle-|VH\rangle)(\langle HV|-\langle VH|)}{2}+(1-p)\frac{I}{4}$. Note that since the coherence length of the photon is only 176 $\lambda$ (due to the 4 nm bandwidth of the interference filter (IF)), much smaller than the optical path difference which is about 0.5 m, two states from the reflected and transmission path only mix at the second BS rather than coherently superpose. In the adaptive measurement module, the two-photon product measurements are realized by the combinations of quarter-wave plates, half-wave plates, polarizing beam splitters, single photon detectors and a coincidence circuit. The rotation angles of quarter-wave plates and half-wave plates can be adaptively adjusted by a controller according to the analysis of the collected coincidence data on a computer.

\begin{figure}[]
\centering
\subfigure[]{
\label{fig2:subfig:a}
\includegraphics[width=0.47\textwidth]{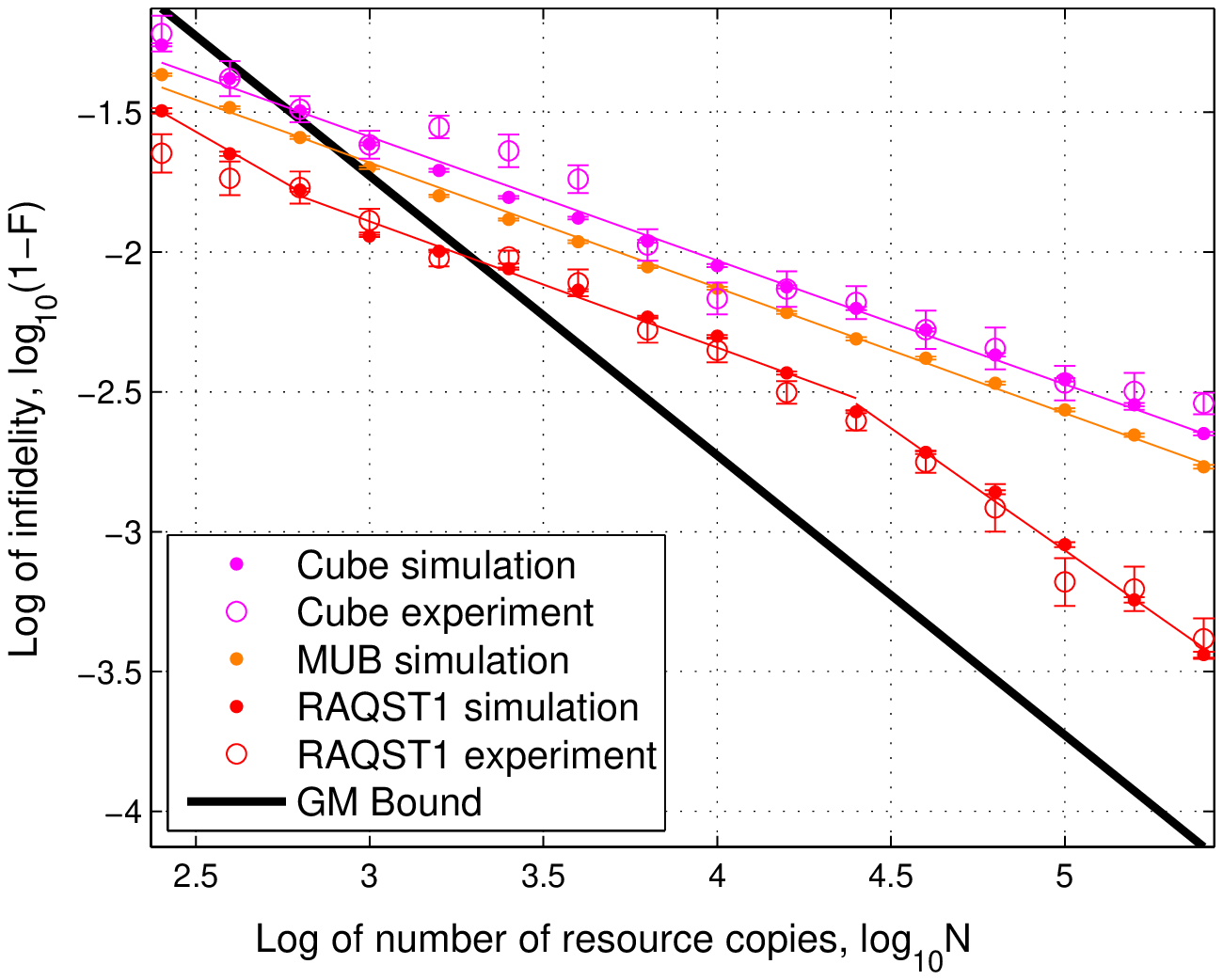}}
\subfigure[]{
\label{fig2:subfig:b}
\includegraphics[width=0.47\textwidth]{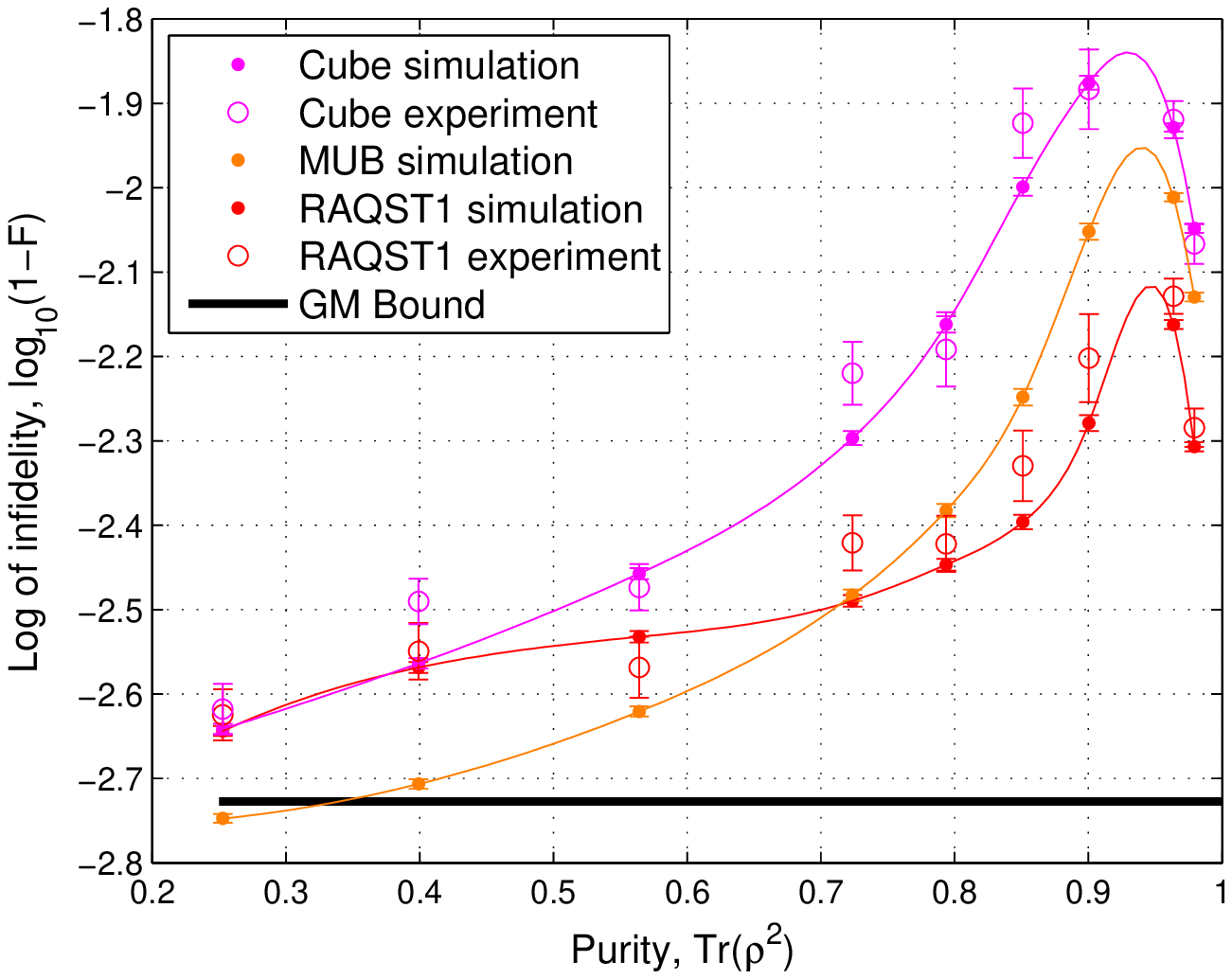}}
\caption{Experimental results. (a) Infidelity versus $N$. Dots are the average infidelity of simulation results with 1000 repetitive runs of RAQST1 (red), MUB (khaki) and standard cube measurements (magenta), and circles are the corresponding average infidelity of 10 repetitive experimental results. Error bars are the standard deviation of the average. (b) Infidelity with respect to purity for $N=10^4$. Simulation (dots) and experimental (circles) results are the average of 1000 and 40 repetitions, respectively.}
\label{fig2:subfig}
\end{figure}

In the first experiment, as shown in Fig.~\ref{fig2:subfig:a}, we realize RAQST1 and standard cube measurements tomography protocols for  entangled states with a high level of purity with respect to different number of resources $N$ ranging from 251 to 251189. First, we calibrate the true state $\rho$ using RAQST1 with $N=10^7$ copies so that the infidelity of the calibrated true state is even 10 times smaller than the estimate accuracy achieved at $N=251189$ with RAQST1. The purity of the calibrated state is 0.983. Systematic error is crucial in the experiments. Beam displacers, which separate extraordinary and ordinary light, act as PBS and have an extinction ratio of about 10000:1. As the precision of rotation stages of QWPs and HWPs are 0.01$^\circ$, the rotation error is determined by the calibration error of optic axes, which is 0.1$^\circ$ in our experiment. Phase errors of the currently used true zero-order QWPs and HWPs are $1.2^\circ$, which dominate the systematic error of practically realized measurements. These error sources induce a systematic error to the estimate state, which can be characterized by its infidelity from the true state. The systematic error is in the order of 10$^{-3}$ when the error sources take the above values. For resource number $N\geq 10^3$, the systematic error is of the same scale as or even larger than the statistical error due to finite resources ($N$ copies). To deal with this problem, we employ error-compensation measurements \cite{Hou15error} to reduce the systematic error to the order of $10^{-5}$. In error-compensation measurement technique, multiple nominally equivalent measurement settings are applied to sub-ensembles such that the systematic errors can cancel out in first order. Tomography experiments using both RAQST1 and standard cube measurements are repeated 10 times for each number of photon resources.
%The reason that there are kinks in the curve corresponding to the RAQST1 is the same as that in Fig.~\ref{fig:subfig:a}.

In the second experiment, as shown in Fig.~\ref{fig2:subfig:b}, we realize tomography protocols using RAQST1 and standard cube measurements for Werner states with purities ranging from 0.25 to 0.98. The purities are changed by adjusting the apertures. Since the photon resource for each run of tomography protocols  is only 10$^4$, we use 10$^6$ copies to calibrate the true state. There are 40 experimental runs and 1000 simulation runs for each of nine Werner states. In each RAQST experiment, four adaptive steps are used to optimize the measurements. To ensure measurement accuracy, error-compensation measurements are also employed.

In both of these two experiments, our experimental results agree well with simulation results.  The improvement of RAQST1 protocol over standard cube measurements strategy is significant. According to the simulation results of MUB protocol and the experimental results of RAQST1, even only with the simplest product measurements, our RAQST1 can outperform the tomography protocols using mutually unbiased bases for states with a high level of purity. Taking into account the trade-off between accuracy and implementation challenge, from Fig.~\ref{fig:subfig} and Fig.~\ref{fig2:subfig}, RAQST using the simplest product measurement seems to be the best choice for reconstructing entangled states with a high level of purity.

%\textbf{Summary and perspectives}
\section{\label{sec:summary}Summary and perspectives}

We presented a new recursively adaptive quantum state tomography protocol using an adaptive LRE algorithm and reported a two-qubit experimental realization of the adaptive tomography protocol. In our RAQST protocol, no prior assumption is made on the state to be reconstructed. The infidelity of the adaptive tomography is greatly reduced and can even beat the Gill-Massar bound by adaptively optimizing the POVMs that are performed at each step.
We demonstrated that the fidelity by using our RAQST with only the simplest product measurements  can even surpass those by using mutually unbiased bases and the two-stage MUB adaptive strategy for states with a high level of purity. Considering the trade-off between accuracy and implementation challenge, it seems that RAQST using the product measurements is the best choice for reconstructing the pure and nearly pure entangled states, which are the most important resources for quantum information processing. It is worth stressing that our RAQST protocol is flexible and extensible. For any finite dimensional quantum systems, once the admissible measurement set is given, we can utilize the adaptive measurement strategy to recursively estimate an unknown quantum state. As demonstrated by  numerical results, if nonlocal measurements can be experimentally realized as the breakthrough of the technology, the admissible measurement set $\mathcal{M}$ can be enlarged, and our RAQST protocol can be better utilized accordingly. How to give a more effective empirical formula in the second stage is worthy of further exploring, where the formula depends upon the estimate state of the first stage. This is actually related to the tomography problem wherein some prior information is already known, e.g., pure entangled states, matrix-product states, low-rank states. By taking full advantage of the prior information, more efficient RAQST protocol may be designed. Our RAQST protocol may have wide applications in practical quantum  tomography experiments.

\emph{Note added}. After we completed the experiments, recently we became aware of a highly relevant work \cite{Stru15experimental} taking a Bayesian estimation approach to realize two-qubit adaptive quantum state tomography using factorized measurements.

\section*{Acknowledgements}
The authors would like to thank Huangjun Zhu for helpful discussions about the Gill-Massar bound for infidelity. The work was supported by the National Natural Science Foundation of China under Grants (Nos.  61222504, 11574291, 61374092 and 61227902) and the Australian Research Council's Discovery Projects funding scheme under Project DP130101658 and Centre of Excellence CE110001027.

%\textbf{Appendix}
\appendix

%\textbf{The recursive ALRE protocol}
\section{\label{sec:append: recursive ALRE protocol}Recursive LRE protocol}

First, we transform the linear regression equations (\ref{average2}) into a compact form. After $t$ times of POVMs, we can obtain in total $M_t=\sum_{k=1}^t M^{(j_k)}$ linear regression equations. We denote them as [1, $(j_1)$], $\cdots$, [$M^{(j_1)}$, $(j_1)$], $\cdots$, [1, $(j_t)$], $\cdots$, [$M^{(j_t)}$, $(j_t)$], where $[m, (j_k)]$ corresponds to the linear regression equation where the outcome $m$ appears when the POVM $\mathcal{M}^{(j_k)}$ is being performed at the $k$-th step. To facilitate the presentation, we rename them according to the natural order. Thus, the notation $\hat{p}(m|\mathcal{M}^{(j_k)})$, $\gamma^{(j_k)}_{m,0}$, $\Gamma^{(j_k)}_m$, $e^{(j_k)}_m$, and $W^{(j_k)}_m$ can be simplified with the corresponding sequence number $n$ as $\hat{p}_n$, $\gamma_{n,0}$, $\Gamma_n$, $e_{n}$, and $W_n$. Let $$Y_t=\left(
          \begin{array}{ccccc}
            \hat{p}_1-\frac{\gamma_{1,0}}{d}, & \cdots, & \hat{p}_n-\frac{\gamma_{n,0}}{d}, & \cdots, & \hat{p}_{M_t}-\frac{\gamma_{{M_t,0}}}{d} \\
          \end{array}
        \right)^{T}
,$$ $$X_t=\left(
          \begin{array}{ccccc}
            \Gamma_1, & \cdots, & \Gamma_n, & \cdots, & \Gamma_{M_t} \\
          \end{array}
        \right)^{T}
,$$ $${\bf e_t}=\left(
          \begin{array}{ccccc}
            e_1, & \cdots, & e_n, & \cdots, & e_{M_t} \\
          \end{array}
        \right)^{T},$$ $$W_t=\textmd{diag}\left(
          \begin{array}{ccccc}
            W_1, & \cdots, & W_n, & \cdots, & W_{M_t} \\
          \end{array}
        \right)^{T}.$$ Using the notation, the linear regression equations (\ref{average2}) can be expressed into a compact form
\begin{equation*}\label{average3}
Y_t=X_t\Theta+{\bf e_t}.
\end{equation*}
The solution of (\ref{theta}) is
\begin{equation}\label{ls}
\hat{\Theta}_{t}=(X_t^{T}W_tX_t)^{-1}X^{T}_tW_tY_t.
\end{equation}

We now show how to recursively obtain the solution of $\hat{\Theta}_t$.
Define
\begin{equation}\label{definitionofq}
Q_n=(\sum^{n}_{i=1}W_{i}\Gamma_{i}{\Gamma_{i}}^{T})^{-1},\ a_n=(\frac{1}{W_{n}}+{\Gamma^{T}_{n}}Q_{n-1}\Gamma_{n})^{-1}.
\end{equation}
Using the matrix inversion formula (see, e.g., page 19 of \cite{Horn12matrix})
$$(A-BCD)^{-1}=A^{-1}+A^{-1}B(C^{-1}-DA^{-1}B)^{-1}DA^{-1},$$
for $n=1, \cdots, M_t$, we have
\begin{equation}\label{recursiveofq}
Q_n=Q_{n-1}-a_nQ_{n-1}\Gamma_{n}{\Gamma^{T}_{n}}Q_{n-1}.
\end{equation}
From (\ref{ls}), (\ref{definitionofq}) and (\ref{recursiveofq}), the
recursive form of $\hat{\Theta}_n$ can be obtained as
\begin{equation}\label{main}
\hat{\Theta}_n=\hat{\Theta}_{n-1}+a_nQ_{n-1}\Gamma_{n}(\hat{p}_n-\frac{\gamma_{n,0}}{d}-{\Gamma^{T}_{n}}
\hat{\Theta}_{n-1}).
\end{equation}

\section{\label{sec:append: mean squared error matrix}Significance of the Mean squared error matrix}
%\textbf{The mean squared error matrix}

As pointed out in \cite{GillM00}, as the number of copies $N$ becomes large, the only relevant measure of the quality of estimation becomes the mean squared error matrix $E(\hat{\Theta}-\Theta)(\hat{\Theta}-\Theta)^{T}\equiv V(\hat{\Theta},\Theta)$. To be specific, for a good estimation strategy, a reasonable expectation is that the elements of the mean squared error matrix decrease as $O(1/N)$, i.e., $E(\hat{\theta}_i-\theta_i)(\hat{\theta}_j-\theta_j)\equiv V_{ij}(\hat{\Theta},\Theta)=O(1/N)$. Assume that $f(\hat{\Theta}, \Theta)$ is any smooth cost function that can measure how much the estimate  $\hat{\Theta}$ ($\hat{\rho}$) differs from the true value $\Theta$ ($\rho$). From equations (1)--(3) in \cite{GillM00}, there exist a function $f_0(\Theta)$ and a positive semidefinite matrix $C(\Theta)$ such that the mean value of $f(\hat{\Theta}, \Theta)$ under a reasonable estimation strategy will decrease as
\begin{equation}\label{cost}
Ef(\hat{\Theta}, \Theta)=f_0(\Theta)+\frac{1}{2}\textmd{Tr}(C(\Theta)V(\hat{\Theta},\Theta))+o(1/N).
\end{equation}
Note that $f_0(\Theta)$ and $C(\Theta)$ depend only on the cost function and the true state, while $V(\hat{\Theta},\Theta)$ depends on the true state as well as the estimation $\hat{\Theta}$. Hence, from (\ref{cost}), we can minimize the mean squared error matrix $V(\hat{\Theta},\Theta)$ to minimize any smooth cost function by choosing appropriate POVMs and suitable estimation strategies.
%This is  possible since there is a partial order for positive semidefinite matrices, namely, $V_1(\Theta)\geq V_2(\Theta)$ means that $V_1(\Theta)-V_2(\Theta)$ is positive semidefinite.
In the following, we only minimize $\textmd{Tr}(V(\hat{\Theta},\Theta))$ instead of $V(\hat{\Theta},\Theta)$ itself for simplicity, although this is not equivalent.

%\textbf{The optimization of POVMs}
\section{\label{sec:append: optimization of POVMs}Optimization of POVMs}

Thanks to the recursive algorithm in Appendix A, we can improve the tomography accuracy by adaptively optimizing the POVMs to be performed in the following step. In order to give a criterion on how to optimize the POVMs, we first look at the mean squared error matrix of $\hat{\Theta}_n$. From (\ref{ls}) and (\ref{definitionofq}), we have
\begin{equation*}\label{variance}
E(\hat{\Theta}_n-\Theta)(\hat{\Theta}_n-\Theta)^{T}=Q_nX_n^{T}W_n\Sigma^{(n)}W_nX_nQ_n,
\end{equation*}
where $\Sigma^{(n)}$ is the covariance matrix of ${\bf e_n}=(e_1, \cdots, e_n)^{T}$.
As we have mentioned in Appendix B, we will minimize the trace of $E(\hat{\Theta}_n-\Theta)(\hat{\Theta}_n-\Theta)^{T}$. This can be achieved by minimizing $Q_n$. Here, we provide an intuitive explanation on why it is reasonable to minimize $Q_n$. It can be seen that if the weighted matrix $W_n$ satisfies $W^{-1}_n\approx\Sigma^{(n)} $  when $N$ becomes large, $E(\hat{\Theta}_n-\Theta)(\hat{\Theta}_n-\Theta)^{T}\approx Q_n$.  Recall that the weight of the $i$-th linear regression equation is approximately equal to the inverse of $\Sigma^{(n)}_{ii}$. Moreover, if $e_i$ and $e_j$  correspond to different POVMs, they are independent, and so $\Sigma^{(n)}_{ij}=0$. Therefore,  we can adaptively choose POVMs to minimize $Q_n$.

Our RAQST protocol is presented as follows. It can be divided into two stages. In the first stage, we  perform a standard linear regression estimation on $N_1$ copies with the standard cube measurement bases to get a prelimiary $\hat{\Theta}$ and $Q$ \cite{Qi13quantum}. Next in the second stage we set the initial value $Q_0=Q$ in (\ref{recursiveofq}) and $\hat{\Theta}_0=\hat{\Theta}$ in (\ref{main}), and then utilize the remaining $N-N_1$ copies for adaptive linear regression estimation.

Suppose after $s$ steps, we get $Q_{M_s}$ and $\hat{\Theta}_{M_s}$ where $M_s=\sum_{k=1}^sM^{(j_k)}$. Recall that $\mathcal{M}^{(j_k)}$ denotes the POVM $\mathcal{M}^{(j_k)}=\{E_m^{(j_k)}\}_{m=1}^{M^{(j_k)}}$ being performed at the $k$-th step. If $s=0$, $M_s=0$. From (\ref{recursiveofq}), we can see that $Q_{M_s+1}\leq Q_{M_s}$, and
\begin{widetext}
\begin{equation}\label{gn}
\textmd{Tr}(Q_{M_s+1})-\textmd{Tr}(Q_{M_s})=-\frac{\Gamma_{M_s+1}^{T}Q^2_{M_s}\Gamma_{M_s+1}}
{\frac{1}{W_{M_s+1}}+\Gamma^{T}_{M_s+1}Q_{M_s}\Gamma_{M_s+1}}\equiv-\textbf{g}_{M_s+1}.
\end{equation}
\end{widetext}
The remaining question is how to choose POVMs to improve the rate of decreasing.
 We can choose $E_i^{(j_{s+1})}$ ($\Gamma_i^{(j_{s+1})}$) from the admissible measurement set
$\mathcal{M}={\underset{j=1}{\bigcup}}\mathcal{M}^{(j)}={\underset{j=1}{\bigcup}}
\{E_i^{(j)}\}^{M^{(j)}}_{i=1}$ such that it maximizes $\textbf{g}_{M_{s}+1}$. Once $E_i^{(j_{s+1})}$ is chosen,
we can perform the corresponding POVM $\mathcal{M}^{(j_{s+1})}=\{E_k^{(j_{s+1})}\}^{M^{(j_{s+1})}}_{k=1}$ at the $(s+1)$-th step. By doing this, we can get $M^{(j_{s+1})}$ linear regression equations. Thus, we can utilize (\ref{recursiveofq}) and (\ref{main}) to get $Q_{M_{s+1}}$ and $\hat{\Theta}_{M_{s+1}}$, where $M_{s+1}=\sum_{k=1}^{s+1}M^{(j_{k})}$. The above procedure is repeated until all the copies have been measured.

Two points should be paid attention to the above RAQST protocol. The first one is when choosing $E_i^{(j_{s+1})}$ to maximize ${\bf g}_{M_s+1}$, we cannot get the information of $\hat{p}_{M_{s}+1}$ in $W_{M_{s}+1}$ because we have not really performed the experiments. From (\ref{average2}), we can use its estimate ${\bf\tilde{p}}_{M_{s}+1}(\hat{\Theta}_{M_s})=\gamma^{(j_{s+1})}_{i,0}/{d}+\hat{\Theta}_{M_s}^T\Gamma^{(j_{s+1})}_i$ to replace $\hat{p}_{M_{s}+1}.$
Another one is that given the total number of copies $N$, how to determine  the number of  copies $N_1$  used in the first stage and the number of adaptive steps in the second stage.
The optimal values remain open. We can give empirical formulas when the dimension of quantum systems for tomography is given.

%\textbf{ 2-qubit ALRE}
\section{\label{sec:append: 2-qubit ALRE}two-qubit RAQST}

In two-qubit RAQST, we assume that the basic admissible measurement set at each iteration step is the standard cube measurement bases, wherein all the elements are one-dimensional projectors. If we choose a projector by minimizing $\textbf{g}$ in (\ref{gn})  as described above, it is easy to reconstruct the corresponding POVM. Actually, if the chosen projector is $|\psi_1\rangle\langle\psi_1|\otimes|\psi_2\rangle\langle\psi_2|$, the corresponding POVM is $\{ |\psi_1\rangle\langle\psi_1|\otimes|\psi_2\rangle\langle\psi_2|, \ |\psi^{\bot}_1\rangle\langle\psi^{\bot}_1|\otimes|\psi_2\rangle\langle\psi_2|,\
|\psi_1\rangle\langle\psi_1|\otimes|\psi^{\bot}_2\rangle\langle\psi^{\bot}_2|,\ |\psi^{\bot}_1\rangle\langle\psi^{\bot}_1|\otimes|\psi^{\bot}_2\rangle\langle\psi^{\bot}_2|\}$, where $|\psi^{\bot}_i\rangle$ is orthogonal to $|\psi_i\rangle$.

As pointed out in \cite{Mahl13adaptive}, in order to minimize infidelity, we must accurately estimate the small eigenvalues of the state to be reconstructed, particularly those nearly 0 eigenvalues. Inspired by this, at each iteration step we find a product projector that minimizes ${\bf\tilde{p}}$.  This projector can be found by a simple iteration algorithm since it is a standard conditional extremum problem. The details are given in Appendix E. By doing this, the value of $\textbf{g}$ may become larger with this new projector. Thus, we add the corresponding POVM into the admissible measurement set at each iteration step in RAQST1. In RAQST2, we further add the set of the eigenbases of the current state estimate into the admissible measurement set. Note that the admissible measurement set in RAQST2 will involve nonlocal measurements in general.

\section{\label{sec:append: Iterative algorithm}Iterative algorithm to minimize ${\bf\tilde{p}}$ within product projectors}

We illustrate our algorithm in the case of two-qubit tomography. We first fix the basis set $\{\Omega_{i}\}^{3}_{i=0}$ for one-qubit, and then we can represent the one-qubit projector measurement for the $j$-th qubit $\{\Pi^j_1, \Pi^j_2\}$ as  $$\Pi^{j}_i=(\pi_{i,0}^{j},\pi_{i,1}^{j},\pi_{i,2}^{j},\pi_{i,3}^{j})^T,\ i,\ j=1,\ 2.$$
Here, the projector $\Pi^j_1$ is orthogonal to $\Pi^j_2$.

For a two-qubit system, we take the tensor product of $\{\Omega_{i}\}^{3}_{i=0}$ as the  basis set. Then any two-qubit product projectors can be parameterized as $$\Pi^1_i\otimes\Pi^2_j=(\sum^{3}_{k=0}\pi_{i,k}^{1}\Omega_k)\otimes(\sum^{3}_{l=0}\pi_{j,l}^{2}\Omega_l).$$ Moreover, we can represent the quantum state as $$\rho=\sum^{3}_{k,j=0}\theta_{4k+j}\Omega_k\otimes\Omega_j.$$ We further introduce the vectorization function $$vec(W_{n\times n})=(W_{11},W_{21},...,W_{n1},W_{12},...,W_{n2},...,W_{nn})^T$$ and assume a $4\times4$ real matrix $P$ such that $$vec(P)=\Theta=(\theta_0,\theta_1,...,\theta_{15})^T.$$

With the notation we can calculate
\begin{equation}
\begin{array}{rl}
{\bf\tilde{p}_{i,j}}&=\text{Tr}(\rho\Pi^1_i\otimes\Pi^2_j)=\Theta^T(\Pi^{1}_i\otimes\Pi^{2}_j)\\
&=vec(P)^Tvec(\Pi^{2}_j{\Pi^{1}_i}^T)\\
&=\text{Tr}(P^T\Pi^{2}_j{\Pi^{1}_i}^T)={\Pi_j^{2}}^TP\Pi_i^{1}.\\
\end{array}
\end{equation}

Usually we  take $\Omega_0=\frac{I}{\sqrt{2}}$. Thus, we have $\pi_{i,0}^{1}=\pi_{j,0}^{2}=\frac{1}{\sqrt{2}}$ and $P_{11}=\frac{1}{2}$. We then denote $\Pi^{1}_i=(\frac{1}{\sqrt{2}},x^T)^T$, $\Pi^{2}_j=(\frac{1}{\sqrt{2}},y^T)^T$ and $P=\left(
\begin{array}{cc}
1/2 & {P_a}^T\\
P_b & P_D\\
\end{array}
\right).$ Now we have ${\bf\tilde{p}_{i,j}}=\frac{1}{4}+\frac{1}{\sqrt{2}}y^TP_b+\frac{1}{\sqrt{2}}P_a^Tx+y^TP_Dx$.

To minimize ${\bf\tilde{p}_{i,j}}$ under the constraints ${\Pi^{1}_i}^T\Pi^{1}_i={\Pi_j^{2}}^T\Pi_j^{2}=1$, we introduce Lagrange multipliers $\lambda_1$, $\lambda_2$, and write $L=\frac{1}{4}+\frac{1}{\sqrt{2}}y^TP_b+\frac{1}{\sqrt{2}}P_a^Tx+y^TP_Dx+\lambda_1(x^Tx-0.5)+\lambda_2(y^Ty-0.5)$. At extreme points we should have
\begin{equation}\label{eqex}\left\{
\begin{array}{rl}
\frac{\partial L}{\partial x}&=\frac{1}{\sqrt{2}}{P_a}+P_D y+2\lambda_1x=0,\\
\frac{\partial L}{\partial y}&=\frac{1}{\sqrt{2}}{P_b}+P_Dx+2\lambda_2y=0.\\
\end{array}\right.
\end{equation}

From (\ref{eqex}) we can design an iterative algorithm: i) choose $x_0=y_0=0$; ii) in step $k$, let $x_k=-\frac{1}{2|\lambda_1|}(\frac{1}{\sqrt{2}}{P_a}+{P_D}^Ty_{k-1})$ and $y_k=-\frac{1}{2|\lambda_2|}(\frac{1}{\sqrt{2}}{P_b}+{P_D}x_{k-1})$, where $\lambda_1$ and $\lambda_2$ are chosen such that ${x_k}^Tx_k={y_k}^Ty_k=0.5$; iii) repeat (ii) until $|L_{k}-L_{k-1}|$ is small enough.

The convergence of our algorithm is straightforward to prove. Note that it can be verified $L_{k}\leq L_{k-1}$ during iterations. Moreover, since ${\bf\tilde{p}_{i,j}}\geq0$, we have $L\geq0$. Therefore, the sequence $\{L_{k}\}$ indeed has a limit and the convergence is guaranteed accordingly.  It is worth noting that $\rho$ is actually unknown. Hence, we should take its current estimate instead of $\rho$ in the above procedure.

\section{\label{sec:append: GM bound}Gill-Massar bound for infidelity in two-qubit state tomography}

According to (A.8) in \cite{Zhu12quantum}, the infidelity between two states $\rho$ and $\hat{\rho}$ is related to the squared Bures distance $D_B^2$ by
\begin{equation}\label{eq:relation infidelity and Bures distance}
1-F(\rho, \hat{\rho})=D_B^2-D_B^4/4.
\end{equation}
Thus, the infidelity and the squared Bures distance share the same Gill-Massar bound in the first order approximation.
Since the Gill-Massar bound for the mean squared Bures distance is $\frac{1}{4}(d+1)^2(d-1)\frac{1}{N}$
in a $d$-dimensional quantum state tomography with a total number of copies $N$ ((5.29) in \cite{Zhu12quantum}), one can derive the Gill-Massar bound $\frac{75}{4N}$ for the infidelity in two-qubit state tomography.


%\bibitem{alre}
%Here and in the following, ALRE means our RAQST protocol but without distinguishing between ALRE1 and ALRE2.
%
%\bibitem{bound}
%This can be gotten by combining the equations (5.29) and (A.8) in \cite{Zhu12quantum}.
%
%Mira14optimal condition number
%Xian11entanglement
%
%\end{thebibliography}

%\textbf{Acknowledgments}
%The authors would like to thank Lei Guo for helpful discussion. The work in USTC is supported by National Fundamental Research Program (Grants No. 2011CBA00200 and No. 2011CB9211200), National Natural Science Foundation of China (Grants No. 61108009 and No. 61222504), Anhui Provincial Natural Science Foundation(No. 1208085QA08). B. Q. acknowledges the support of National Natural Science Foundation of China (Grants No. 61004049, No. 61227902, No. 61374092 and No. 61134008). D. D. is supported by the Australian Research Council (DP130101658).
%
%\textbf{Author contributions}
%B.Q., Y.W. and D.D. developed the scheme based on linear regression model, Z.-B.H., G.-Y.X and G.-C.G performed the numerical simulations. All authors discussed the results and contributed to the writing of the paper. G.-Y.X supervise the project.
%
%\textbf{Additional information}
%
%\textbf{Competing financial interests:} The authors declare no competing financial interests.

\end{document}